\documentclass[authoryear,11pt]{elsarticle}
\usepackage{graphicx}
\usepackage{amsmath,amsfonts}
\usepackage{lineno}
\usepackage{subfigure}
\usepackage[colorlinks]{hyperref}
\newcommand{\ud}{\mathrm{d}}

\begin{document}
\begin{frontmatter}
\title{Fluctuation Domains in Adaptive Evolution}
\author[carl]{Carl Boettiger\corref{cor1}}
\ead{cboettig@ucdavis.edu}
\cortext[cor1]{Corresponding author.}
\author[dushoff]{Jonathan Dushoff\corref{cor1}}
\ead{dushoff@mcmaster.ca}
\author[weitz,w2]{Joshua S. Weitz\corref{cor1}}
 \ead{jsweitz@gatech.edu}

\address[carl]{Center for Population Biology, University of California, Davis, United States}

\address[dushoff]{Department of Biology, McMaster University,
Hamilton, ON, Canada}
\address[weitz]{School of Biology, Georgia Institute of Technology,
Atlanta, United States}
\address[w2]{School of Physics, Georgia Institute of Technology,
Atlanta, United States}
\begin{abstract}
We derive an expression for the variation between parallel
trajectories in phenotypic evolution, extending the well known result that
predicts the mean evolutionary path in adaptive dynamics or
quantitative genetics.
We show how this expression gives rise to the notion of fluctuation domains --
parts of the fitness landscape where the rate of evolution is very predictable
(due to \emph{fluctuation dissipation}) and parts where it is highly
variable (due to \emph{fluctuation enhancement}).  These fluctuation
domains are
determined by the curvature of the fitness landscape.  Regions of the fitness
landscape with positive curvature, such as adaptive valleys or
branching points,
experience enhancement. Regions with negative curvature, such as adaptive
peaks, experience dissipation.  We explore these dynamics in the ecological
scenarios of implicit and explicit competition for a limiting resource.
\end{abstract}
\end{frontmatter}
\section{Introduction}
Fitness landscapes have long been an important metaphor in evolution.
\citet{Wright31} originally introduced the concept to explain his
result that the mean rate of evolution of a quantitative trait is
proportional to gradient of its fitness.  The result and its
accompanying metaphor continue to arise in evolutionary theory, having
been derived independently in quantitative genetics~\citep{Lande79},
game-theoretic dynamics~\citep{hofbauer_book1998, Abrams93}, and
adaptive dynamics~\citep{dieckmann_jmb1996}.  The metaphor creates a
deterministic image of evolution as the slow and steady process
of hill climbing.  Other descriptions of evolution have focused on its
more stochastic elements -- the random chance events of mutations and
the drift of births and deaths that underlie the
process~\citep{kimura_book1984, kimura1968, ohta_pnas2002}.  In this
manuscript, we seek to characterize the deviations or fluctuations of
evolutionary trajectories
around the expected evolutionary path.

Our main result is that the size of deviations of evolutionary
trajectories
is determined largely by the \emph{curvature} of the evolutionary landscape
whose \emph{gradient} is determining the mean rate of evolution.
The landscape metaphor can be used to understand the interplay of
stochastic and deterministic forces by identifying regimes where the selection
will counterbalance or enhance such stochastic fluctuations.
Because curvature can be positive (concave up) or negative (concave down), the
evolutionary landscape can be divided into domains where fluctuations are
enhanced or dissipated.  To make this more precise, we focus on a
Markov model of
evolution used in the theory of adaptive dynamics.  This Markov model gives
rise to the familiar gradient equation -- like that which first inspired the
fitness landscapes metaphor -- and also a more precise statement of the
fluctuation dynamics.  The adaptive dynamics framework is general enough that
we can consider how these results apply in a variety of ecological scenarios.
We illustrate the surprising consequence of bimodal distributions of expected
phenotypes among parallel trajectories emerging from a fixed starting point
on a single-peak adaptive landscape due to fluctuation enhancement.
We also provide a landscape interpretation that suggests how the ideas
of fluctuation dynamics apply to other ecological and evolutionary
models.

\section{Theoretical construction}
Various definitions of fitness landscape have been used in
the study of evolutionary dynamics.
The conventional idea postulates a mapping between trait values and
fitness~\citep{levins62, levins64, rueffler04}.  The difficulty with
this conception is that, in
many cases, fitness of a given population type
is mediated largely through competition with
other populations, and is thus dependent on the current distribution
of populations
and their respective phenotypes in the environment.
This dependence on densities or frequencies turns the static
fitness landscape into a dynamic landscape,
whose shape changes as the number of individuals, and their
corresponding traits, changes.

It is possible to depict the fitness landscape that emerges even in the
face of density- or frequency-dependent
competition by returning to Wright's notion of a
landscape. To do so, consider
a resident population of individuals in an environment, each of which have
an identical trait value, \emph{i.e.}, a monomorphic population.
Next, consider the per-capita fitness of a small number of mutant individuals
with a similar, but different trait.  The per-capita fitness of mutants
in the environment may be larger, smaller or identical to the per-capita
fitness of residents.
The gradient in fitness around the trait represented by the resident
is termed the selective derivative~\citep{geritz_prl1997} in adaptive
dynamics (analogous to the selective gradient in quantitative
genetics).
The fitness landscape is defined by integrating over these derivatives in trait
space, to obtain a \emph{local} picture of how fitness changes.  In
the limit of
small, slow mutation, it can be shown that the population will evolve
as predicted
by the shape of the landscape: by climbing towards a peak most rapidly when the
slope is steep~\citep{rueffler04}, see Fig.~\ref{1} for three
illustrative examples.   The trait on the horizontal axis is that of the
resident, not the mutant.  The lower panels show the slope of the mutant
fitness as a function of the resident trait (up to a multiplicative factor)-- the slope of the mutant fitness
changes as the resident trait changes. The trait of the resident
population evolves in the direction given by that slope.  Integrating
the lower panel along the resident trait axis reveals the fitness
landscape
the resident trait climbs in an evolutionary process (top panels).
Note that changes in the mutant fitness are captured in the changes in
slope of the landscape; the fitness landscape itself remains fixed.
We derive the expected evolutionary dynamics of the phenotypic trait of the
resident and its variation
among parallel trajectories in the following sections.

\subsection{Model}
Consider a population monomorphic for a particular phenotypic trait,
$x$.  The abundance of the population, $N(x,t)$ of trait $x$,
is governed by its ecological dynamics:
\begin{equation}
\frac{\mathrm{d} N(x)}{\mathrm{d} t}=f(x,N,E), \label{eco}
\end{equation}
which may depend on the trait, population abundance, and the
environmental conditions, $E$.  The evolutionary dynamics proceed in
three steps ~\citep{dieckmann_jmb1996, champagnat2006}.  First, the
population assumes its equilibrium-level abundance $N^{\ast}(x)$,
determined by its phenotypic trait $x$.  Next, mutants occur in the
population at a rate given by the individual mutation rate $\mu$ times
the rate of births at equilibrium, $b(x)$.  The mutant phenotype, $y$,
is determined by a mutational kernel, $M(x,y)$.
The success of the mutant strategy depends on the invasion fitness,
$s(y,x)$, defined as
the per-capita growth rate of a rare population.
Mutants with negative invasion fitness die off, while mutants with
positive invasion
fitness will invade with a probability that depends on this
fitness~\citep{geritz_prl1997}.

The invasion fitness is calculated from the ecological dynamics,
Eq.~\eqref{eco}, and will in general depend on the trait $x$ of the
resident population as well as that of the mutant, $y$.  We assume
that a successful invasion results in replacement of the
resident~\citep{geritz_prl1997, geritz_jmb2002}, and the population
becomes monomorphically type $y$.  The details of this formulation can
be found in Appendix~\ref{AdaptiveDynamics}.  The important
observation is that such a model can be represented by a Markov
process on the space of possible traits, $x$.  The population can jump
from any position $x$ to any other position $y$ at a transition rate
$w(y|x)$ determined solely by the trait values $x$ and $y$.  The
probability that the process is at state $x$ at time $t$ then obeys
the master equation for the Markov process,

\begin{equation}
\frac{\mathrm{d}}{\mathrm{d} t} P(x, t) = \int \mathrm{d} y\  \left[
w(x|y)P(y, t) - w(y|x) P(x,t)\right].
\label{evo.eq.master}
\end{equation}

No general solution to the master equation~\eqref{evo.eq.master} exists.
However, if the jumps from state $x$ to state $y$ are sufficiently small
(\emph{i.e.}, if mutants are always close to the resident), we can obtain approximate
solutions for $P(x,t)$ by applying a method known as the Linear Noise
Approximation~\citep{vankampen_book2001},
Appendix~\ref{LinearNoiseApproximation}.  Doing so yields
a general solution for the probability $P(x,t)$ in terms of the transition
rates, $w(y|x)$, Eq.~\eqref{FP}.  The linear noise approximation
\emph{derives} a diffusion equation as an approximation
to the original jump process, as is commonly postulated.  While this diffusion
can be written as a partial differential equation (PDE), following Kimura, or
as a stochastic differential equation, the solution for the probability density
can be proven to be Gaussian and hence it suffices to write down ordinary
differential equations for the first two moments~\citep{kurtz_1971}.

\subsection{The Fluctuation Equation}
 
We assume the mutational kernel $M(y,x)$ is Gaussian in the difference between 
resident and mutant traits, $y-x$, with width $\sigma_{\mu}$ and that mutations 
occur at rate $\mu$.   A more thorough discussion of these quantities can be 
found in Appendix~\ref{AdaptiveDynamics}, where they are developed in the process 
of deriving the transition probability $w(y|x)$ that specifies the underlying 
Markov process.  Using Eq.~\eqref{FP} which results from the systematic expansion 
of the master equation~\eqref{evo.eq.master}, the mean trait $\hat x$ obeys 
\begin{equation}
\frac{\mathrm{d} \hat x}{\mathrm{d} t} = \frac{1}{2}\mu \sigma_{\mu}^2 N^*(\hat x) \partial_y s(y, \hat x)|_{y=\hat x} \equiv a_{1}(\hat x), \label{explicitCanonical}
\end{equation}
with an expected variance $\sigma^2$ that obeys 
\begin{equation}
\frac{\partial \sigma^2}{\partial t} = 2 \sigma^2 \partial_{\hat x} a_{1}(\hat x) + 2\sqrt{\tfrac{2}{\pi} }\sigma_{\mu} \left|a_1(\hat x) \right| \label{explicitFluctuation} .
\end{equation}
 
Eq.~\eqref{explicitCanonical} recovers the familiar canonical equation of adaptive 
dynamics~\citep{dieckmann_jmb1996} for the mean trait.   
Eq.~\eqref{explicitFluctuation}, which describes the variance, we term the 
\emph{fluctuation equation} of adaptive dynamics.    The linear
noise approximation (see Appendix~\ref{LinearNoiseApproximation}) also 
predicts that the probability distribution itself is Gaussian, so that 
Eqs.~\eqref{explicitCanonical} and~\eqref{explicitFluctuation}
determine the entire distribution of possible trajectories in this approximation.  To illustrate the local behavior of the model, we define the fitness
landscape as $L(x) = \int_0^x a_1(y)\,dy$.

 
\begin{figure}
\begin{center}
	\subfigure[Implicit Resource]{\label{Logistic}\includegraphics[width=2in]{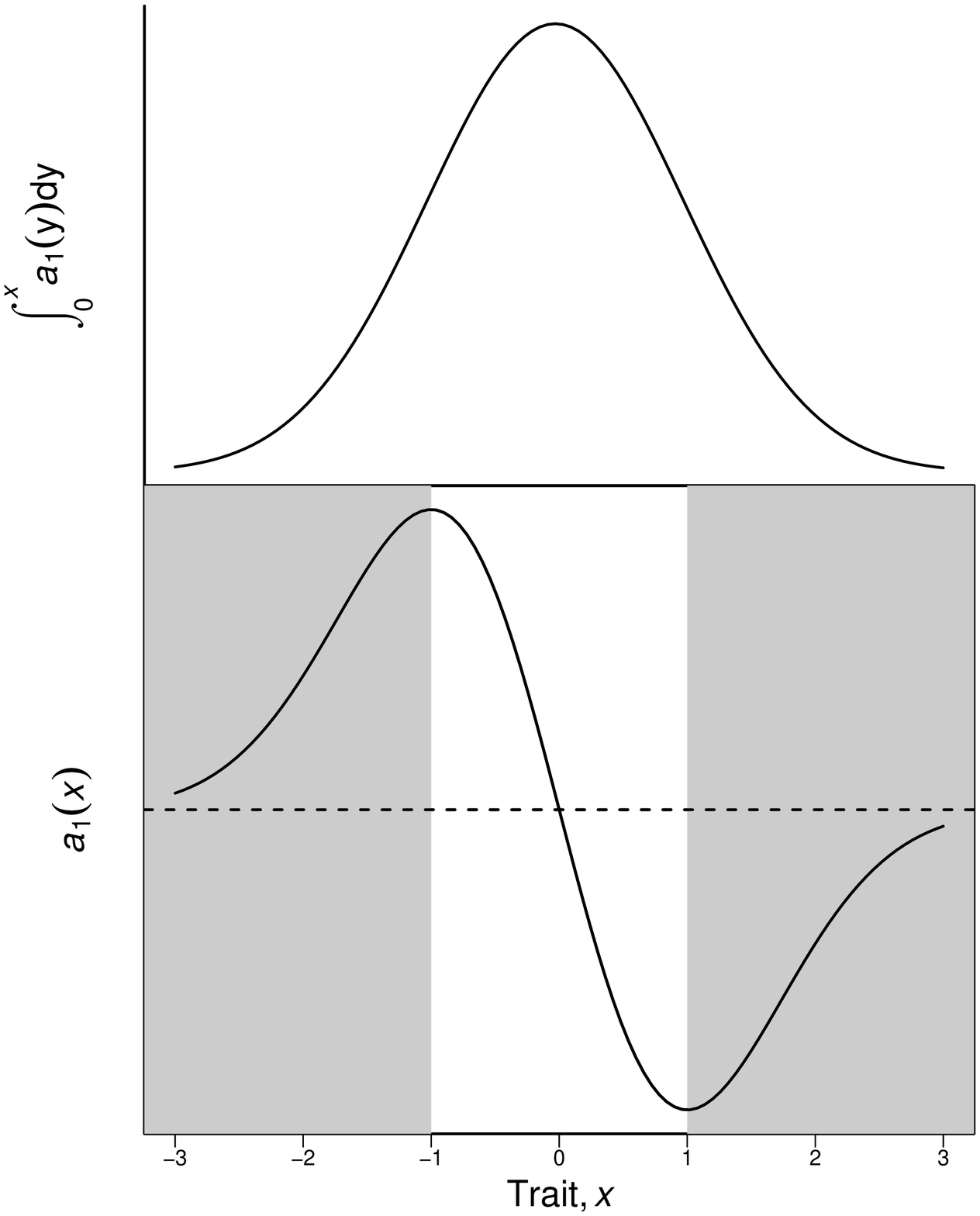}}\subfigure[Chemostat]{\label{Chemostat}\includegraphics[width=2in]{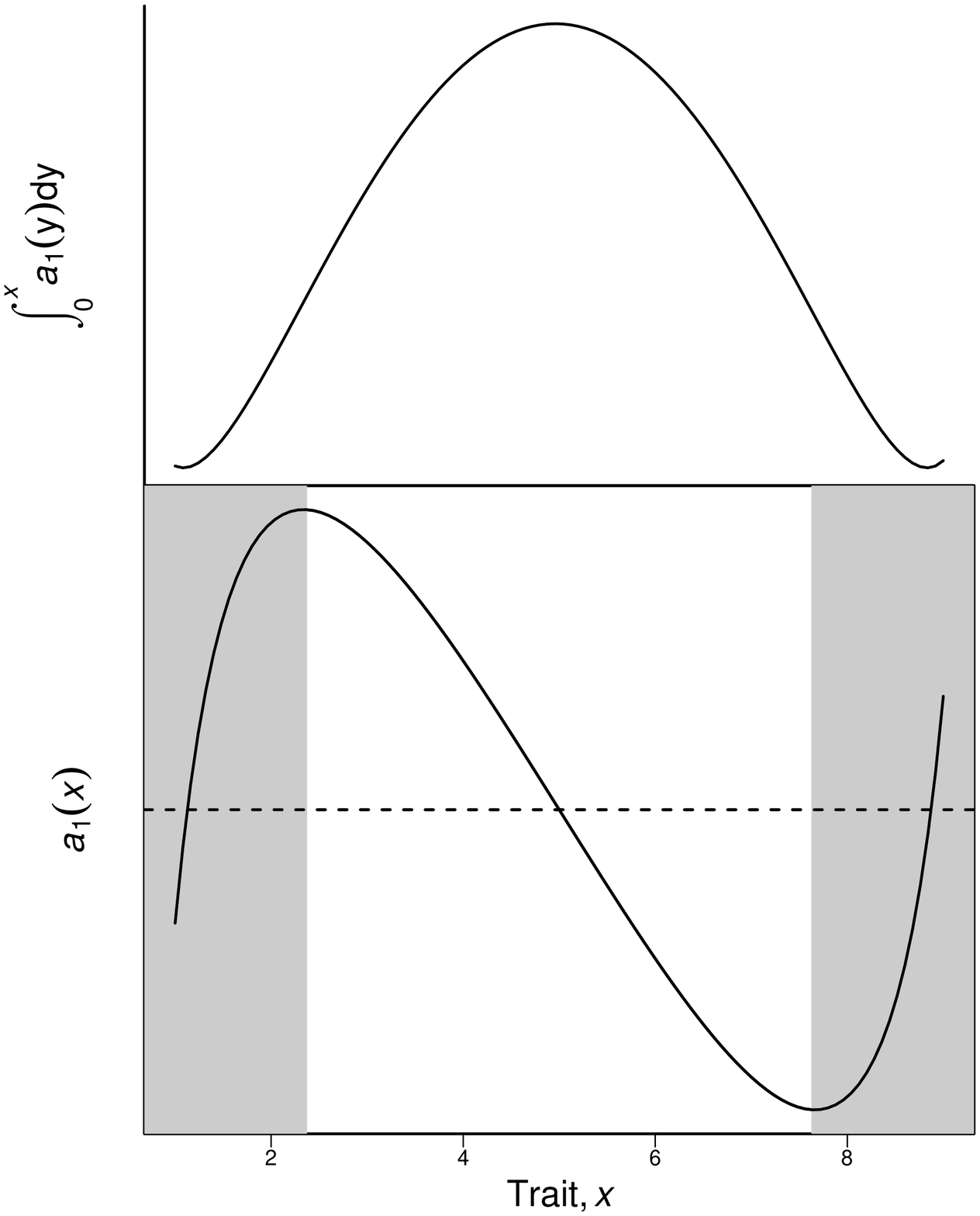}}\subfigure[Branching]{\label{Branching}\includegraphics[width=2in]{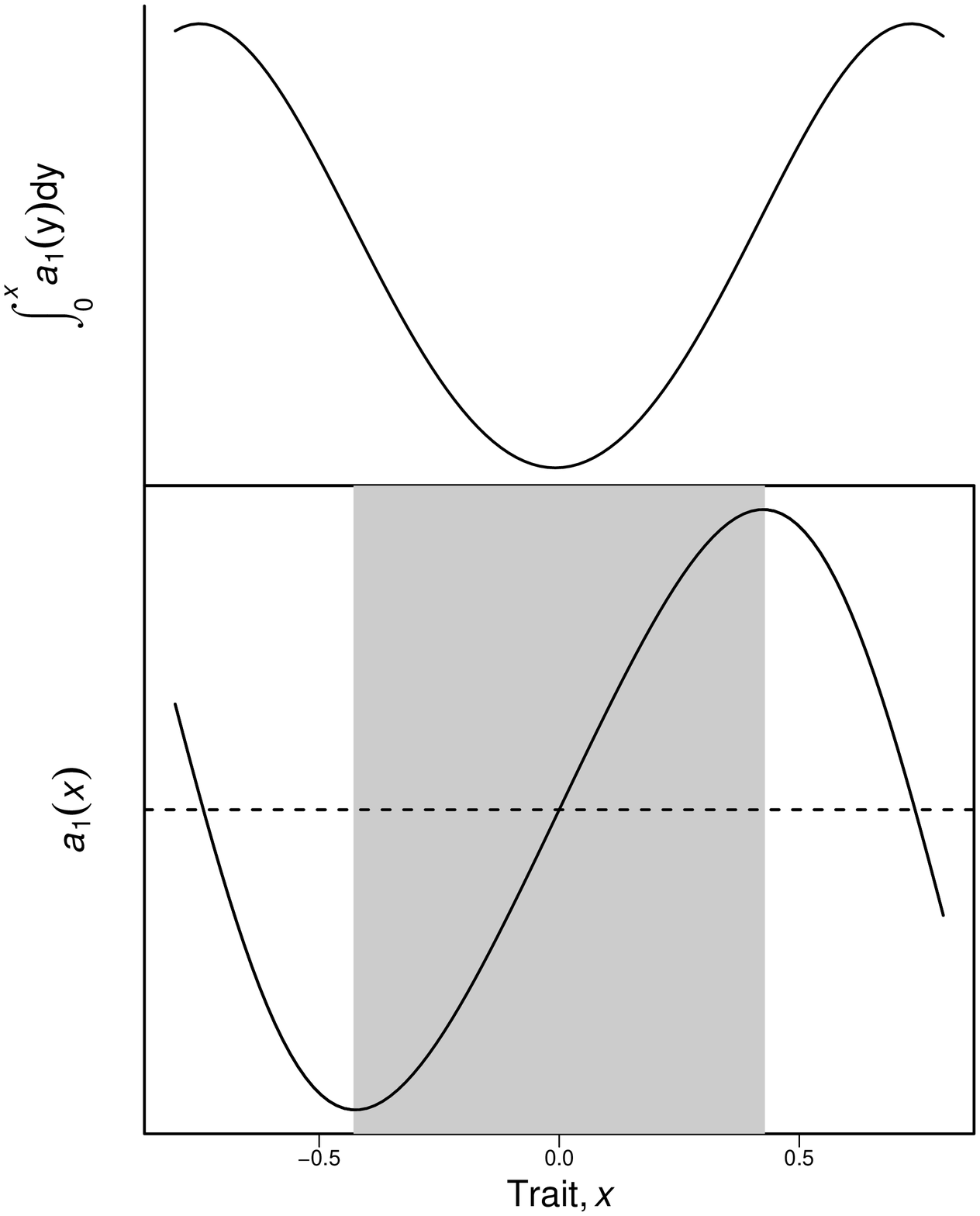}}
 
	\caption{\textbf{Fluctuation domains}.   The lower panels show the rate of 
evolution given by Eq.~\eqref{explicitCanonical} vs. the resident trait $x$ for three ecological models: 
(a) implicit competition for a limiting resource (Section~\ref{compete}), 
(b) explicit resource competition~(Appendix~\ref{ChemostatModel}), and 
(c) symmetric branching for dimorphic populations (with a resident population of trait $x$ and another at $-x$, Appendix~\ref{Equations}).  The horizontal dashed lines in the lower panel denote where $a_1(x)=0$, \emph{i.e.}, the selective derivative is zero, and 
correspond to evolutionarily singular points, \emph{e.g.}, adaptive peaks and valleys.
Shaded and unshaded regions in the lower panels correspond to trait values for
which fluctuations are enhanced (shaded) and dissipated (unshaded).
The adaptive landscape is defined as the integral of this expression (upper panels) -- where populations climb the hills at a rate proportional to their steepness.  Note that
the adaptive landscape for the symmetric branching case has two peaks, and so
evolutionary trajectories will lead to a stable dimorphism.}\label{1}
\end{center}
\end{figure}

\section{Fluctuation Domains}
 
A closer look at Eq.~\eqref{explicitFluctuation} will help motivate our geometric interpretation of fluctuation domains.  The approximation requires that the mutational step size $\sigma_{\mu}$ is very small, hence the second term will almost always be much smaller than the first (except when the fluctuations $\sigma$ themselves are also very small). Note that the second term resembles the right-hand side of~\eqref{explicitCanonical}, differing only by a small scalar constant and the fact that it is always positive.  Interestingly, this means that the second term vanishes at the singular point where $a_1(\hat x) = 0$ and $\sigma^2=0$: fluctuations cannot be introduced at an evolutionary equilibrium.
That the first term is the gradient of the deterministic (mean) dynamics and the second term is smaller by a factor of the small parameter in the expansion are general features of the linear noise approximation.  
 
The first term of~\eqref{explicitFluctuation} implies that the variance $\sigma^2$ will increase or decrease exponentially at a rate determined by $\partial_x a_1(x)$; \emph{i.e.}, the \emph{curvature} of the fitness landscape.  This landscape and its gradient are depicted in Fig.~\ref{1} for several ecological scenarios.   Plotting the gradient itself makes it easier to see when fluctuations will increase or decrease.  On the gradient plot the singular point is found where the curve crosses the horizontal axis.
 
In the neighborhood of a singular point corresponding to an adaptive
peak, the slope is negative, hence the first term in Eq.~\eqref{explicitFluctuation} (the coefficient of $\sigma^2$) is negative and fluctuations dissipate exponentially.  This means that two populations starting with nearby trait values will converge to the same trajectory.  In this region no path will stochastically drift far from the mean trajectory, hence the canonical equation will provide a good approximation of all observed paths.  We term the part of the trait-space landscape where $ \partial_x a_1(x) < 0$ the fluctuation-dissipation domain, analogous to the fluctuation-dissipation theorem found in other contexts~\citep{vankampen_book2001}.  
 
Farther from the singular point corresponding to an adaptive peak,
$\partial_x a_1(x)$ becomes positive (see Fig.~\ref{Logistic} and Fig.~\ref{Chemostat}).  While this part of trait-space still falls within the basin of attraction of the singular point, the variance $\sigma^2$ between evolutionary paths will grow exponentially.  Initially identical populations starting with trait values in this region will experience divergent trajectories due to this enhancement.  The variation can become quite large, with evolutionary trajectories that differ significantly from the mean.  We call this region of trait-space where $\partial_x a_1(x) > 0$ the fluctuation enhancement domain.  Eventually trajectories starting in this region will be carried into the fluctuation-dissipation domain, where they will once again converge.  

Evolutionary branching points~\citep{dieckmann_nat1999, geritz_evoeco1998} provide another example of a fluctuation enhancement domain. Until now, we have assumed that successful mutants replace the resident population, a result known as ``invasion implies substitution''~\citep{geritz_jmb2002}. However, at an evolutionary branching point, the mutant invader no longer replaces the resident population, and the population becomes dimorphic.  In dealing with monomorphic populations, we have been able to describe the evolutionary dynamics in terms of the change of a single trait, describing a single resident population.  For dimorphic populations, two resident populations coexist, and both influence the shape of the landscape. Thus, an invader's fitness, $s(y;x_1,x_2)$, and its rate of evolution, $a_1(x_1,x_2)$, will depend on how it performs against both populations.  The evolutionary dynamics after branching are two-dimensional and require a multivariate version of Eq.~\eqref{explicitFluctuation}. Despite this, we can still gain qualitative insight using the intuition that connects fluctuation domains to curvature.

The existence of the stable dimorphism fundamentally distorts the landscape.  While the population was monomorphic, being closer to the singular point always ensured a higher fitness. Once a resident population sits on either side of a branching point though, the singular point becomes a fitness \emph{minimum}. This description of evolution towards points which become fitness minima after branching has been addressed extensively elsewhere~\citep{geritz_prl1997, geritz_evoeco1998, geritz_jmb2004}. Here, what interests us is the effect of branching on fluctuations. If the landscape is smooth, a minimum must have positive curvature and therefore must be an enhancement domain. The farther a mutant gets from the branching point, the faster it can continue to move away, thus enhancing the initially small differences between mutational jumps. A rigorous description of this effect would require a multivariate version of Eq.~\eqref{explicitFluctuation} which is beyond the scope of this paper.  Instead, we illustrate the enhancement effect by taking
a slice of the two-dimensional landscape by assuming that the dimorphic populations have symmetric trait values about the singular strategy, $x_1=-x_2=x$. The fitness landscape for the case of symmetric branching is shown in Fig~\ref{Branching}, and the derivation provided in Appendix~\ref{BranchingModel}. Note that the region around the branching point is a fluctuation enhancement regime and the two new fitness maxima far from the branching point are in fluctuation dissipation regimes.

To provide a more concrete illustration of the fluctuation dynamics, we will focus on the description of the ecological competition model and compare the theoretical predictions of Eq.~\eqref{explicitCanonical} and Eq.~\eqref{explicitFluctuation} to point-process simulations of the Markov process~\citep{Gillespie}.

\section{An Example of Fluctuation Domains in Resource Competition}
\label{compete}
\subsection{An Ecological Model of Implicit Competition for a Limiting Resource}
The logistic model of growth and competition in which populations
compete for a limited resource is a standard model in ecological dynamics~\citep{dieckmann_nat1999}.  Here, we consider
the population dynamics of $N(x,t)$ individuals each with trait $x$:
\begin{equation}
\frac{\mathrm{d} N(x,t)}{\mathrm{d} t} = rN(x,t)\left(1-\frac{\sum_yN(y,t)C(x,y)}{K(x)}\right),\label{general_logistic}
\end{equation}
where $r$ is the birth rate and $\sum_yrN(y)C(x,y)/K(x)$ is the density-dependent death rate.  
In the model, $K(x)$ is the equilibrium population density and $C(x,y)$ is a function which describes the relative change in death rate of individuals of type $x$ due to competition by individuals of type $y$.  Given $C(x,x)=1$, the equilibrium density of a monomorphic population is $N^{\ast}(x)=K(x)$.  Following~\citet{dieckmann_nat1999} we assume the following Gaussian forms,
\begin{equation}
K(x)=K_0e^{-x^2/(2\sigma_k^2)}, \label{K}
\end{equation}
and
\begin{equation}
C(x,y)=e^{-(x-y)^2/(2\sigma_c^2)}, \label{C}
\end{equation}
where $\sigma_k$ and $\sigma_c$ are scale factors for the resource distribution and competition kernels respectively.  We focus on the case $\sigma_c>\sigma_k$, for which the model has a convergence-stable evolutionarily stable strategy (ESS) at $x=0$~\citep{geritz_evoeco1998}.  Having specified the ecological dynamics, we must also specify the evolutionary dynamics, which we assume occur on a much slower time scale. We assume that with probability $\mu$ an individual birth results in a mutant offspring, and that the mutant trait is chosen from a Gaussian distribution centered at the trait value of the parents and with a variance $\sigma_{\mu}^2$.  From this we can assemble $w(y|x)$, and compute equations~\eqref{explicitCanonical} and~\eqref{explicitFluctuation} (see Appendix~\ref{LinearNoiseApproximation}).  
 
\subsection{Numerical simulation of stochastic evolutionary trajectories}
There are many ways to represent evolutionary ecology models in numerical simulations, including finite simulations, fast-slow dynamics, and point processes~\citep{dieckmann_nat1999,nowak_science2004,champagnat2006,champagnat_2007}.  We choose to model the evolutionary trajectories of the ecological model in Eq.~\eqref{general_logistic} by explicitly assuming a separation of ecological and evolutionary time scales.  First, given a trait $x$, the ecological equilibrium density $N^{\ast}(x)$ is determined.  Then, the time of the next mutant introduction is calculated based on a Poisson arrival rate of $\mu r N^{\ast}(x)$.  In essence, we are assuming that ecological dynamics occur instantaneously when compared to evolutionary dynamics.  The trait of the mutant, $y$, is equal to $x$ plus a normal deviate with variance $\sigma_{\mu}^2$.  
The mutant replaces the resident with probability given by the standard branching process result, $1-d(y,x)/b(y,x)$ where $d$ and $b$ are the per-capita birth and death rate of the 
mutant, respectively, so long as $b>d$, and with probability 0 otherwise.  
The ecological steady state is recalculated and the process continues.  
Simulations are done using Gillespie's minimal process method~\citep{Gillespie}.
Estimates of the mean phenotypic trait $\hat x$ and the variance $\sigma^2$ are calculated from ensemble averages of at least $10^5$ replicates. 
 
\begin{figure}
\begin{center}
\includegraphics[width=.9\textwidth]{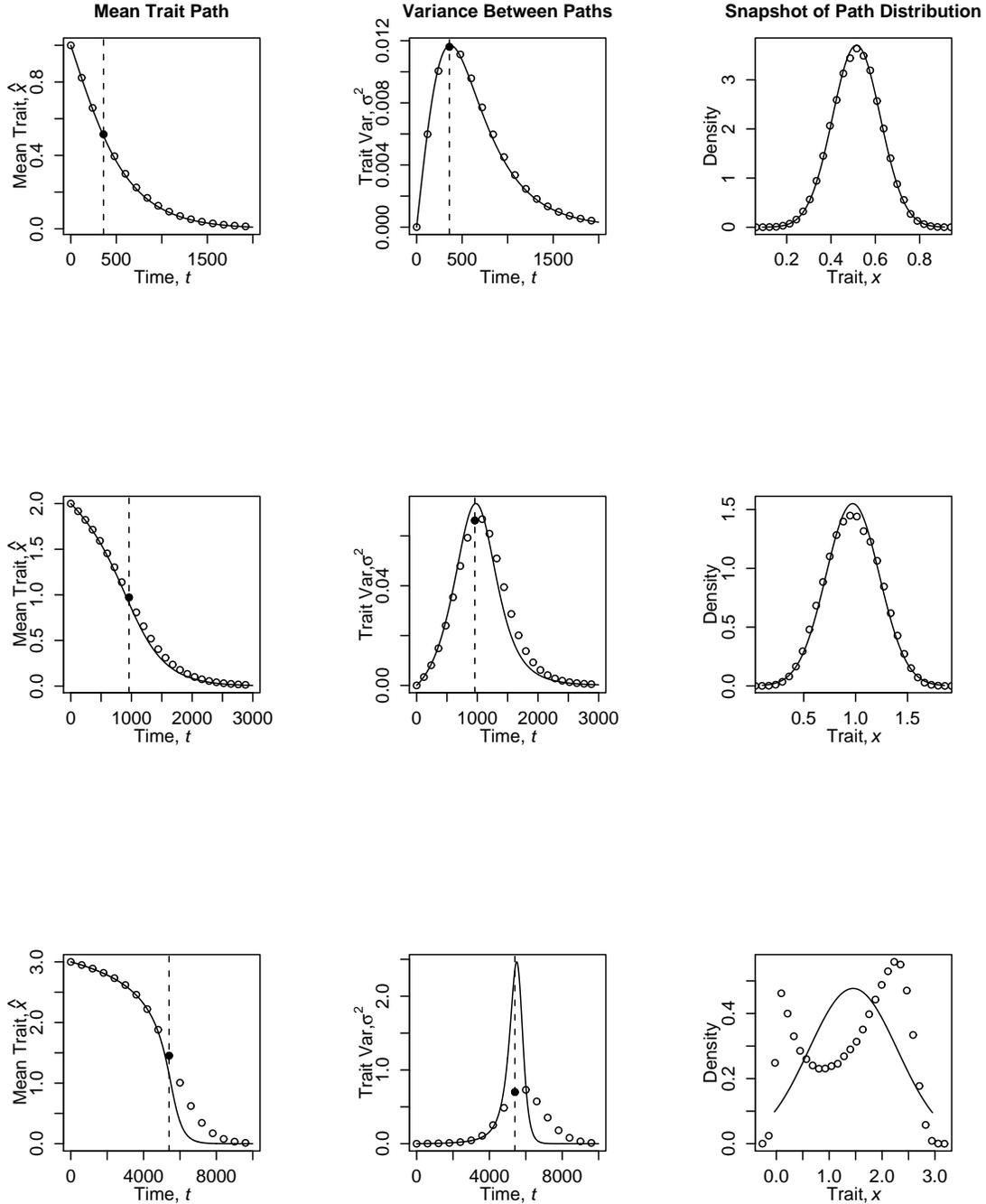}
\caption{\textbf{Simulations over fluctuation domains}.  The dynamics of the mean path, variance between paths, and snapshots of the trait distribution in time for three initial conditions: $x_0=1$, $x_0=2$, and $x_0=3$ (top to bottom successively).  The three rows correspond to trajectories experiencing fluctuation dissipation (tow) and fluctuation enhancement (bottom) along with an intermediate case (middle). The mean (first column) and variance (second column) among trait values are plotted over time.  Circles are simulation averages from $10^5$ replicates and lines are theoretical predictions.  The final column shows the theoretical Gaussian distribution (solid line) at the point in time indicated by the dashed line
and closed circle in the first two columns, and the 
actual histogram (circles) of positions across the replicates at that time.  
The bottom right panel shows a bimodal distribution among replicates in the case
of fluctuation enhancement.  
\textbf{Parameters:} $\sigma_k^2 = 1$, $\sigma_c^2 = 1.01$, $r=10$, $\sigma_{\mu} = 0.0005$, $\mu = 1$.}\label{2}
\end{center}
\end{figure}
 
\subsection{Comparison of Theory and Simulation}
We consider simulations with three different initial conditions: starting within the fluctuation-dissipation domain ($x = 1)$, just into the fluctuation-enhancement domain ($x = 2$), and deep into the fluctuation enhancement regime ($x = 3)$ in successive rows in Fig.~\ref{2}.  For each condition, the simulation is repeated $10^5$ times and we compare the resulting distribution of trajectories to that predicted by theory.  In the first column we plot the numerically
calculated mean path taken to the ESS (shown in circles), and 
compare to the theoretical prediction of Eq.~\eqref{explicitCanonical}.  In the second column, we plot the variance between paths, and compare to the predictions of Eq.~\eqref{explicitFluctuation}.  As each replicate begins under identical conditions, the initial variance is always zero. In the third column, we present the distribution of traits among the replicates at a single instant in time.  We chose the moment of maximum variance so that deviations from the theoretically predicted Gaussian kernel will be most evident. 
 
\subsubsection{Fluctuation-Dissipation Domain}
 Starting at the edge of the dissipation domain, fluctuations grow for only a short time before rapidly decaying, in the first row of Fig.~\ref{2}.  Since the simulation is sampled at regular time intervals, the vertical spacing of points indicate the rate of evolution.  Theoretical predictions closely match simulation results for both the mean and variance, and the resulting distribution matches the theoretically predicted Gaussian. The variance represents an almost negligible deviation from the mean trajectory.   
 
\subsubsection{Fluctuation-Enhancement Domain}
Our second ensemble at $x=2$ begins clearly within the fluctuation-enhancement domain.  In the second row of Fig.~\ref{2}, the variance plot (center panel) begins by growing exponentially.  Once the population crosses into the dissipation domain, at $x=1$, at about $t=1000$, the variance peaks and then dissipates exponentially.  This transition appears as an inflection point in the mean path (left panel).  Though the variation now represents significant deviations, the distribution still appears Gaussian (right panel).  
 
\subsubsection{Strong Fluctuations}
In the third row of Fig.~\ref{2}, starting deep within the fluctuation-enhancement domain at $x = 3$, theory and simulation agree initially.  Remaining long enough in this domain, fluctuations are eventually on the same order of magnitude as the macroscopic dynamics themselves, and the linear noise approximation underlying the theory begins to break down.  The theory and simulation variance diverge, while the mean path is substantially slower than predicted by the canonical equation.   In the right panel, the reason for this becomes clear.  At this point in the divergence, the distribution is far from Gaussian, rather it has become bimodal, with some replicates having reached the stable strategy at $x=0$ and others having hardly left the initial state.  
 
The mechanism behind the emergence of bimdal probability distributions for trajectories can be seen in the final row of Fig.~\ref{2}.  The theoretical trajectory predicted by Eq.~\eqref{explicitCanonical} is sigmoidal, comprised of slow change at the beginning and end and rapid evolution in the middle.  Ecologically, this arises at the beginning from the small population size available to generate mutations, and at the end from the weakening selection gradient.  Populations with intermediate trait values hence experience rapid trait evolution, giving rise to this transiently bimodal distribution.  The theory provides insight into this case of strong fluctuations in two ways -- first, it is driven by the existence of a fluctuation-enhancement domain, and second, it is anticipated by the theoretical prediction of fluctuations that are on the same order as the macroscopic dynamics.  We observe that these macroscopic fluctuations can arise whenever the population remains long enough in the enhancement regime, independent of other model details.  The fluctuation equation for the explicit resource competition (chemostat) model shown in Fig.~\ref{Chemostat} and described in Appendix~\ref{ChemostatModel} never predicts fluctuations that reach macroscopic values, and simulations of this system (not shown) do not produce bimodal trait distributions.  
 
\section{Discussion}
 
The metaphor of fitness landscapes has been a powerful tool for understanding evolutionary processes~\citep{wright1932,Lande79,Abrams93,dieckmann_jmb1996}.  The concept of a landscape arises naturally from describing the rate of change of the mean trait as being proportional to the evolutionary gradient.  We extend this metaphor to understand fluctuations away from this mean, where our analysis of a Markov process representation of evolution in the adaptive dynamics framework brings us back to the notion of a fitness landscape.  In this case, it is the curvature of that landscape which informs the dynamics of these fluctuations away from the expected evolutionary path.  Our result linking landscape curvature to fluctuation domains is consistent with similar results from quantitative genetics that relate the sign and magnitude of the quadratic selection coefficient in the Price equation to increases or decreases in trait variation~\citep{lande_1983,chevin_gen2008}.
 
Though the mathematical analysis of the Markov process is somewhat technical, the result is generally intuitive.  Graphs such as Fig.~\ref{1} provide a visualization of how fluctuations will behave on the landscape, while our fluctuation equation, Eq.~\eqref{explicitFluctuation}, provides a more explicit description of those fluctuations.  This result, like the gradient expression itself, relies on the underlying randomness being small relative to the scale of evolutionary change.  We have seen how these approximations can break down in the case of very large fluctuations, resulting in a bimodal distribution of phenotypes.  While Eq.~\eqref{explicitFluctuation} fails to describe this case accurately, it predicts the breakdown of the approximation by the explosion of fluctuations.  Though we have focused on the adaptive dynamics model of evolution, we hope that the approach taken here can be extended to other landscape representations.  Further, we hope that our model will be broadly applicable to assessing the repeatability of evolution in experimental studies of microbes and viruses~\citep{wichman_sci1999,cooper_pnas2003} and in ecological studies of rapid phenotypic change caused by biotic interactions~\citep{duffy_ecolett2007} and/or anthropogenic factors (such as over-fishing) \citep{olsen_2005}.
 
In this work, we have compared theoretical predictions
to numerical simulations.  Even when a closed form expression such as~\eqref{explicitFluctuation} exists for the deviations, some of the most dramatic examples in Fig.~\ref{2} emerge only when the diffusion approximation breaks down and stochastic forces become macroscopic.  In the spirit of scientifically reproducible research~\citep{TempleLang07, Claerbout, Stodden}, we freely provide all the source code required to replicate the simulations and figures shown in the text.  Though the numerical simulations are written in C for computational efficiency, we provide a user interface and documentation by releasing all the code, figures, text, and examples as a software package for the widely used and freely available R statistical computing language.

\section{Acknowledgements}
We thank Sebastian Schreiber for many helpful discussions, and also thank Michael Turelli, G\'eza Mesz\'ena and an anonymous reviewer for comments.  This work is funded in part by the Defense Advanced Research Projects Agency under grant HR0011-05-1-0057.  Joshua S.~Weitz, Ph.D., holds a Career Award at the Scientific Interface from the Burroughs Wellcome Fund.  Carl Boettiger is supported by a Computational Sciences Graduate Fellowship from the Department of Energy under grant number DE-FG02-97ER25308.  
 
\appendix
 
\section{Adaptive Dynamics and the Transition Probability $w(y|x)$}\label{AdaptiveDynamics}
 
In this appendix we construct the Markov process $w(y|x)$ under the assumptions of adaptive dynamics~\citep{dieckmann_jmb1996}.  The probability per unit time of making the transition in trait space from a monomorphic population with trait $x$ to one with trait $y$ is given by
\begin{equation}
	w(y|x)=\mathcal{M}(y,x)\mathcal{D}(y,x).
\end{equation}
In the framework presented here, a monomorphic population of residents  with trait $x$ generate mutants with trait $y$, some of which survive.
The rate at which a mutation is generated from a population is 
\begin{equation}
	\mathcal{M}(y,x) = \mu(x) b(x) N^{\ast}(x)M(x, y),
\end{equation}
where $b(x)$ is the per-capita birth rate at equilibrium, $\mu(x)$ the mutation probability per birth, $N^{\ast}(x)$ the equilibrium population size for a population with trait $x$, and $M(x, y)$ is the distribution from which the mutant trait is drawn.  The probability of surviving accidental extinction of a branching process given the mean individual birth rate $b$ and mean death rate $d$ for the mutant $y$ is $\mathcal{D}(y,x)=1-d(y,x)/b(y,x)$ if $d(y,x)<b(y,x)$ and $\mathcal{D}(y,x)=0$ otherwise~\citep{feller1968}.  The terms $b(y,x)$ and $d(y,x)$ refer to the birth and death rate, respectively, of a rare mutant with trait $y$ in an equilibrium population of $x$.  
 
Given a mutant strategy $y$ such that $\mathcal{D}(y,x)>0$ we have
\begin{equation}
	\label{solved_w}
	w(y|x) = \mu(x) N^{\ast}(x) b(x) M(x, y)[b(y,x) - d(y,x)]/b(y,x).
\end{equation}
Expanding the fitness, $b(y,x)-d(y,x)$, to first order 
the transition rate is then
\begin{equation}
	\label{approx_w}
	w(y|x) \approx \mu(x) N^{\ast}(x) \partial_y s(y,x)|_{y=x} M(x, y)[y-x] ,
\end{equation}
where $\partial_y s(y,x)|_{y=x}$ is known as the selective derivative~\citep{geritz_prl1997}.  From Eq.~\eqref{approx_w} one can apply a particular model by specifying expressions for the mutation rate $\mu(x)$, stationary population size, $N^{\ast}(x)$, fitness function $s(y,x)$ and mutational kernel $M(x,y)$.  In the competition for a limited resource model,~\citep{dieckmann_nat1999} used here, these are:
 
\begin{align}
	\mu(x) &= \mu, \nonumber \\
	M(y,x) &= \tfrac{1}{\sqrt{2\pi\sigma_{\mu}^2}}e^{-\frac{(y-x)^2}{2\sigma_{\mu}^2}}, \nonumber \\ 
	s(y,x) &= r\left(1-\frac{N^{\ast}(y)  e^{-\frac{(x-y)^2}{2\sigma_c^2}}  }{N^{\ast}(x)}\right), \nonumber \\
	N^{\ast}(x) &= K_0e^{-\frac{x^2}{2\sigma_k^2}}. \label{explicitS}
\end{align}
Consequently, the evolutionary transition rates in for this model are given by
\begin{equation}
	w(y|x) = -\mu  K_0e^{-\frac{x^2}{2\sigma_k^2}} \frac{rx}{\sigma_k^2} \frac{e^{-\frac{(y-x)^2}{2\sigma_{\mu}^2}} }{\sqrt{2\pi\sigma_{\mu}^2}}[y-x].
	\label{example_w}
\end{equation}
 
The transition rate $w(y|x)$ for the explicit resource competition model is presented along with the model details in Appendix~\ref{ChemostatModel}.  Using the appropriate transition rate in the linear noise approximation described in Appendix~\ref{LinearNoiseApproximation}, we recover the equations for the curves plotted in Fig.~\ref{1} which are integrated to obtain the theoretical predictions of Fig.~\ref{2}.  These explicit expressions are given in Appendix~\ref{Equations}. 
 
\section{Linear Noise Approximation}\label{LinearNoiseApproximation}
\subsection{About the approximation}
The linear noise approximation is a common approach for describing Markov processes.  Though often applied in discrete cases such as one-step (birth-death) processes, it can be generalized to the continuous case we consider, where a population at trait $x$ can jump to another trait value $y$. The approximation transforms the Markov process specified by a master equation on the transition rates $w(y|x)$ to an approximate partial differential equation (PDE) for the probability distribution.  This PDE resembles the Fokker-Planck equation for the process\footnote{Indeed, they are equivalent if transition rates are linear -- in which case the PDE is also exact.}, except that the PDE resulting from the linear noise approximation is guaranteed to be linear and its solutions Gaussian.  Consequently, solving for the two moments, the mean and variance, will lead to a system of ordinary differential equations.  Substituting the form of $w(y|x)$ found in Appendix~\ref{AdaptiveDynamics} into this ODE system recovers Eq.~\eqref{explicitCanonical} and~\eqref{explicitFluctuation} in the text.   
 
The approximation is straight-forward (involving a change of variables and a Taylor expansion), if cumbersome.  The approximation is rigorously justified over any fixed time interval $T$ in the limit of small step sizes~\citep{kurtz_1971}, which parallels more modern justification of the Canonical equation~\citep{champagnat_2001}.  The original derivation of the canonical equation makes use of (unscaled) jump moments, introduced by \citet{vankampen_book2001}.  We review this approach first, as it provides a good intuition for the full linear noise approximation.  The actual approximation relies on a change of variables which makes explicit use of the small step sizes, and \emph{derives} rather than assumes the Gaussian character of the distribution.  
\subsection{Original jump moments}
The dynamics of the average phenotypic trait are given by
\begin{equation}
	\frac{\mathrm{d} \hat x(t)}{\mathrm{d} t} = \int \mathrm{d} x \phantom \cdot x \frac{\mathrm{d}}{\mathrm{d} t} P(x,t).
\end{equation}
Using the master equation
\begin{equation}
\frac{\mathrm{d}}{\mathrm{d} t} P(x, t) = \int \mathrm{d} y\  \left[ w(x|y)P(y, t) - w(y|x) P(x,t)\right],
\end{equation}
to replace $\tfrac{\mathrm{d}}{\mathrm{d} t} P(x,t)$ and performing a change of variables, we find,
\begin{equation}
	\frac{\mathrm{d} \hat x(t)}{\mathrm{d} t} = \int \mathrm{d} x \int \mathrm{d} y [y-x] w(y|x)P(x,t).
\end{equation}
Defining the $k$th jump moment as $a_k = \int (y-x)^k w(y|x) \mathrm{d} y$, the dynamics can be written as,
\begin{equation}
	\frac{\mathrm{d} \hat x(t)}{\mathrm{d} t} = \langle a_1(x) \rangle . \label{exact1}
\end{equation}
It is by no means obvious if or when the deterministic path approximation $\langle a_1(x) \rangle \approx a_1(\langle x \rangle )$ is valid, as $a_1$ will often be nonlinear.  The justification lies in the linear noise approximation.  Proceeding as above, we also find an expression for the second moment,  
 
\begin{equation}
	\frac{\mathrm{d} \langle x^2(t) \rangle }{\mathrm{d} t} = 2\langle x a_1(x) \rangle + \langle a_2(x) \rangle \label{exact2} . 
\end{equation}
Which again, we will only be able to solve by means of the linear noise approximation.   
\subsection{The linear noise approximation}
To justify this step we will change into variables where we can have an explicit parameter $\varepsilon$ that relates to the step size. The trait $x$ is approximated by an average or macroscopic value $\phi$ and a deviation $\xi$ that scales with the mutational step size $\varepsilon$; $x = \phi + \varepsilon \xi$.  Defining $r \equiv y-x$ expand the transition rate $w(y|x)$ in powers of $\varepsilon$, 
 
\begin{equation}
w(y|x) = f(\varepsilon) \left[ \Phi_0(\varepsilon x; r ) +  \varepsilon \Phi_1(\varepsilon x; r ) + \varepsilon^2 \Phi_2+\ldots \right],
\label{transformed_w}
\end{equation}
where the $\Phi$ terms in the expansion are functions in which $\varepsilon$ appears only in terms of $\varepsilon x$.  The function $f(\varepsilon)$ indicates that we can rescale the entire process by some arbitrary factor of $\varepsilon$ since it can always be absorbed into the timescale.  We can then define the transformed jump moments as moments of $\Phi$ rather than $w$,
 
\begin{equation}
\alpha_{\nu, \lambda}(X) = \int r^{\nu} \Phi_{\lambda}(X,r) \mathrm{d} r . 
\end{equation}
 
The probability $P(x,t)$ is expressed in terms of the new variables $P(\phi(t) + \varepsilon \xi, t) = \Pi(\xi,t)$, and the master equation~\eqref{evo.eq.master} becomes:
 
\begin{align}
\frac{\partial }{\partial \tau}\Pi(\xi, \tau) & - \varepsilon^{-1} \frac{\mathrm{d} \phi}{\mathrm{d} \tau} \frac{\partial }{\partial \xi}\Pi(\xi, \tau) \nonumber \\
=& - \varepsilon^{-1} \frac{\partial}{\partial \xi} \alpha_{1,0}(\phi(\tau) + \varepsilon \xi) \cdot \Pi(\xi, \tau) \nonumber \\
+& \frac{1}{2} \frac{\partial^2}{\partial \xi^2} \alpha_{2,0}(\phi(\tau) + \varepsilon \xi) \cdot \Pi(\xi, \tau)  \nonumber \\
-& \frac{1}{3!} \varepsilon \frac{\partial^3}{\partial \xi^3} \alpha_{3,0}(\phi(\tau) + \varepsilon \xi) \cdot \Pi(\xi, \tau) \nonumber \\ 
+& \varepsilon \frac{\partial}{\partial \xi} \alpha_{1,1}(\phi(\tau) + \varepsilon \xi) \cdot \Pi(\xi, \tau)  +\mathcal{O}(\varepsilon^2), 
\end{align}
where we have rescaled time by $\varepsilon^2 f(\varepsilon)t = \tau$. Expanding the jump moments around the macroscopic variable $\phi$, 
\begin{equation*}
\alpha_{\nu, \lambda}(\phi(t)+\varepsilon \xi) \approx \alpha_{\nu, \lambda}(\phi) + \varepsilon \xi \alpha_{\nu,\lambda}' + \frac{1}{2} \varepsilon^2 \xi^2 \alpha_{\nu, \lambda}(\phi)'' + \mathcal{O} \varepsilon^3,
\end{equation*}
(where primes indicate derivatives with respect to $\phi$), and collecting terms of leading order in $\varepsilon$ we have:
\begin{equation}
\frac{\mathrm{d} \phi}{\mathrm{d} \tau} = \alpha_{1,0}(\phi), \label{canonical}
\end{equation}
which is a completely deterministic expression.  Substituting the form of $w(y|x)$ from~\eqref{approx_w} recovers the canonical equation of adaptive dynamics, Eq.~\eqref{explicitCanonical}.  Observe that the fluctuations are an order $\varepsilon$ smaller, demonstrating that this is indeed a consistent approximation.  Collecting terms of order $\varepsilon^0$ we have the partial differential equation
\begin{equation}
\frac{\partial}{\partial \tau}\Pi(\xi, \tau)= -\alpha_{1,0}'(\phi) \frac{\partial}{\partial \xi} \xi \Pi + \frac{1}{2} \alpha_{2,0}(\phi) \frac{\partial^2}{\partial \xi^2} \Pi \label{FP},
\end{equation}
while all other terms are order $\varepsilon$ or smaller.  This is a partial differential equation for the evolution of the probability distribution of traits.  It is a linear Fokker-Planck equation, hence its solution is Gaussian and $\Pi(\xi, t)$ can be described to this order of accuracy by its first two moments, 
\begin{align*}
\frac{\partial \langle \xi \rangle}{\partial t} &= \alpha_{1,0}'(\phi)\langle \xi \rangle, \\
\frac{\partial \langle \xi^2 \rangle }{\partial t} &= 2\alpha_{1,0}'(\phi)\langle \xi^2 \rangle + \alpha_{2,0}(\phi),
\end{align*}
where prime indicates derivative with respect to the trait $x$. If $\alpha_{1,0}'(\phi)<0$ or the initial fluctuations $\langle \xi \rangle_0$ are zero, the first moment can be ignored, and the variance of the ensemble is given by transforming back into the original variables:
\begin{equation}
\varepsilon^2 \frac{\partial \sigma^2 }{\partial t} = 2\alpha_{1,0}'(\phi)\sigma^2 + \alpha_{2,0}(\phi). \label{fluctuation}
\end{equation}
 
Transforming between the scaled variables and the original variables requires the appropriate choice of $\varepsilon$.  The assumption that mutational steps are small provides a natural choice: $\varepsilon = \sigma_{\mu}$.   Substituting Eq.~\eqref{approx_w} to compute the jump moments, this recovers the fluctuation expression~\eqref{explicitFluctuation} in the text.  Note that even before we perform this substitution that~\eqref{fluctuation} has the same form as~\eqref{explicitFluctuation}; fluctuations grow or diminish at a rate determined by the sign of the gradient of the deterministic equation, Eq.~\eqref{canonical}.

\section{Chemostat Model}\label{ChemostatModel}
The graphical model for a second scenario is also displayed in Fig.~\ref{1}. This scenario describes competition and evolution with explicit resource dynamics for a chemostat system.   We consider a resource $Q$ that flows in at rate $D$ from a reservoir fixed at concentration $Q_0$.  To retain constant volume in the chemostat, both biotic and biotic components are flushed from the system at constant rate $D$.  The chemostat contains populations with $N_i$ organisms, each of which take up nutrients at a rate $g_i$ and convert these into reproductive output with efficiency $\eta_i$, 
\begin{align}
	\dot Q & =  D Q_0 - D Q - \mathbf{g}(Q) \mathbf{N}, \\
	\dot N_i & = -D N_i + \eta_i g_i(Q) N_i. 
\end{align}
 
Assume that the uptake of nutrients is governed by Michaelis-Mentin dynamics, and also that a trade-off exists between efficiency of nutrient take-up and conversion (imagining greater investment in foraging means less energy available for reproduction),
\begin{align*}
	g(Q) & = Q/(1+hQ),  \\
	\eta(x) & = x, \\
	h(x) & = x^2,
\end{align*}
where the trait $x$ may differ between populations.
 
The associated equilibrium population size for a population with trait $x$ is 
 
\begin{align*}
	\bar N(x) & = \frac{D Q_0 - D\bar Q}{g(\bar Q, x)} = x Q_0 - \frac{xD}{x-x^2D}.
\end{align*}
Similarly, the resource uptake of a mutant with trait $y$ in an environment at an equilibrium set by a resident type with trait $x$ is given by 
\begin{align*}
	g_y(\bar Q_x) = \frac{1}{x/D-x^2+y^2},
\end{align*}
from which we find the invasion fitness function and its gradient,
\begin{align*}
	s(y,x) &= \frac{y}{x/D-x^2+y^2} - D, \\
	\partial_y s(y,x)|_{y=x} &= \frac{D}{x} - 2D^2.
\end{align*}
 
Assuming a Gaussian mutation kernel and constant mutation rate, from Eq.~\eqref{approx_w} the transition rate function $w(y|x)$ is
\begin{equation}
	w(y|x) \approx \mu \left[ x Q_0 - \frac{xD}{x-x^2D}\right] \left[ \frac{D}{x} - 2D^2 \right]\frac{e^{-\frac{(y-x)^2}{2\sigma_{\mu}^2}} }{\sqrt{2\pi\sigma_{\mu}^2}}[y-x]. 
\end{equation}
 
\section{Branching Model}\label{BranchingModel}
The model of implicit competition for a limiting resource, Eqs.~\eqref{general_logistic}-\eqref{C}, is well known to exhibit the phenomenon of evolutionary branching when the competition kernel is narrower than the resource distribution, $\sigma_c < \sigma_k$~\citep{dieckmann_nat1999}.
Once branching occurs, the invasion fitness of a rare mutant is no longer given by $s(y,x)$ as in Eq.~\eqref{explicitS}, but instead depends on the trait values of each of the coexisting residents $x_1$ and $x_2$, as in $s(y, x_1, x_2)$. 
This invasion fitness can still be calculated directly from the competition model,  Eqs.~\eqref{general_logistic}-\eqref{C}.  This mutant can either arise from the $x_1$ or $x_2$ population and replace it.  If we assume $x_1 = -x_2 = x$, we have the case of a symmetrically branching population.  While many realizations of branching may be close to symmetric, this is but a one-dimensional slice through a two-dimensional trait space $(x_1, x_2)$.   In this case, we can express the equilibrium density of each resident species by $K_{\text res}(x) = K(x)/(1+C(x,-x) )$.  We then consider the initial per capita growth rate of a rare mutant with trait $y$:

\begin{equation}
s(y,x, -x) = r\left( 1 - \frac{K_{\text res}(x)C(x,y) + K_{\text res}(-x) C(-x, y)}{K(y)}\right).
\label{syxx}
\end{equation}
This replaces the $s(y,x)$ function for the monomorphic population, and we proceed as before to calculate $a_1(x)$ in Eq.~\eqref{explicitCanonical}.  That is, we evaluate $\partial_y s(y,x,-x)$ at $y=x$, take $K_{\text res}(x) \equiv N^{\ast}(x)$ to recover a closed-form slice of the branching landscape that is depicted in Fig.~\ref{Branching}.  The expression itself is given in the Appendix~\ref{Equations}, Eq.~\eqref{dimorphic_a1}.

\section{Explicit solutions for examples}\label{Equations}
For the example logistic competition in a monomorphic population, the evolution of the mean trait $x$ is given by:
\begin{equation}
\frac{\ud x}{\ud t} = -\frac{x}{2\sigma^2_k} r \mu \sigma^2_{\mu} K_0 e^{-x/2\sigma^2_k} \label{logistic_a1}.
\end{equation}
In Fig.~\ref{1}, we choose parameters such that $r \mu K_0 \sigma_{\mu}^2 / 2  =1$ and $\sigma_k^2 = 1$.

For the chemostat:
\begin{equation}
	\frac{\ud x}{\ud t} = \frac{1}{2} \mu \sigma^2_{\mu} \left(Qx-\frac{D}{1-xD}\right) (D/x-2D^2) \label{chemostat_a1} .
\end{equation}
In Fig.~\ref{1}, we choose parameters such that $\mu \sigma^2_{\mu} = 1$, $D = 0.1$, and $Q=0.1$.  

And for the symmetrically dimorphic population,
\begin{equation}
	\frac{\ud x}{\ud t} = \frac{-r \mu \sigma^2_{\mu} K_0 	e^{-\tfrac{x^2}{2}\left(\tfrac{4}{\sigma_c^2}-\tfrac{1}{\sigma_k^2}\right) } (\sigma^2_c+e^{\tfrac{2 x^2}{\sigma^2_c}}\sigma^2_c-2\sigma^2_k)x} { \left(1+e^{2x^2/\sigma^2_c}\right)^2 \sigma^2_c \sigma^2_k} \label{dimorphic_a1}.
\end{equation}
The parameters for the dimorphic population plot of Fig.~\ref{1} are chosen such that $r \mu K_0 \sigma_{\mu}^2 / 4 = 1$, $\sigma_k^2 = 2$ and $\sigma_c^2 = 1$.  

The right-hand side of each of these expressions is the function $a_1(x)$ from the text, Eq.~\eqref{explicitCanonical}.  This function also determines the fluctuation dynamics by way of Eq.~\eqref{explicitFluctuation}.  The respective plots of $a_1(x)$ are used in Fig.~\ref{1}, while solving the ODEs for $\hat x(t)$ and $\sigma^2(t)$ gives the theoretical predictions of Fig.~\ref{2}.

\bibliography{mycanonical.bib}
 
\end{document}